\documentclass[a4paper,10pt]{article}

\usepackage[dvips]{graphicx}
\usepackage{epsfig}
\usepackage{graphicx}
\def\be{\begin{equation}}

\def\ee{\end{equation}}
\def\es{\end{split}}
\def\bi{\bibitem}
\usepackage{amsmath}
\usepackage{textcomp}
\usepackage{cite}
\begin{document}
\title{ Wormhole inducing exponential expansion in $R^2$ gravity}
\author{B Modak$^{}$\footnote{E-mail: bijanmodak@yahoo.co.in}  and  {Gargi Biswas$^{}$\footnote{E-mail: biswasgb@gmail.com}  }}

\maketitle
\noindent
\begin{center}
\noindent
Department of Physics, University of Kalyani, Kalyani,India-741235\\

\end{center}

\begin{abstract}
Wormholes are considered both from the Wheeler deWitt equation, as well as from the field equations in the Euclidean background of Roberson Walker mini-superspace in $R^2$ gravity.
Quantum wormhole satisfies Hawking Page wormhole boundary condition in the Euclidean background of mini-superspace, however, in the Lorentzian background wave functional turns to the usual oscillatory function. The Euclidean field equations for $\kappa=0, \pm1$
lead to the wormhole configuration; as well as oscillating universe in Euclidean time $\tau$. The oscillating universe is equivalent to expanding universe under analytic continuation $\tau=it$ and asymptotically leads to exponential solution. An Euclidean wormhole in the very early era evolves to an oscillating universe in $\tau$, thereafter crossing deSitter radius transition to an inflationary era is evident at later epoch only for $\kappa=1$.
 \end{abstract}

\section{Introduction}
Inflationary expansion  is necessary \cite{gu:prd,li:pl,
stei:prl,gu:prl,a:plb,inf6} in the early vacuum dominated universe to resolve some of the problems of standard cosmology, like horizon problem, structure formation problem, etc. However, the initial singularity problem prevails in the cosmological model before the Planck era of strong curvature regime. In the strong curvature regime quantum gravity is necessary, however it revealed that Einstein's gravity is not renormalizable.
\par
In this endeavour the quadratic curvature correction was introduced from the fundamental of Physics to improve the renormalizability of general relativity \cite{stell:prd}. Subsequently, other higher order curvature terms appear naturally as renormalization terms in quantum theory in curved space time \cite{qft:bd}. In this venture modified theory of gravity \cite{capp:prd,harko:prd,lobo:fR} appears as an alternative theory of gravitation. Starobinsky \cite{inf6} first proposed inflationary model based on $R^2$ term. The modified theory of gravitation can accommodate not only inflation, but it can explain late time acceleration of the universe \cite{capgara:cqg,capfran:grg,bm:aspace}. On due course of development low energy effective Heterotic string theory \cite{heter:string,met:string}, Kaluza Klein gravity \cite{sahdev:plb,wsson:AJ,Dzhu:prd}, etc., appear as an alternative theory of gravitation.
\par
The problem of cosmological singularity prevails irrespective of the success of aforesaid gravity theories, whose resolution has not yet been settled. A few proposals have been advocated to alleviate the singularity problem \cite{Kanti:prd, Anto:np}.
First of all, the wave function of the universe \cite{HH:prd} in quantum cosmology is finite even at classical singularity. Further, the Euclidean wormhole that appears in the context of quantum gravity, can alleviate cosmic singularity problem by introducing tunnelling of the universe through the classical singularity. Hawking and Page \cite{hawk:page} proposed that the wave function is finite even at classical singularity and die away at large three space for the existence of quantum wormhole.
\par
Wormhole solutions \cite{morris1,morris2,
lee:prl,ho:plb,hall:cqg,
cou:cqg,lobo:bd,nandi:bd} and their consequences \cite{haw:prd,
co:npb,co:npb1,fisu:plb,
pol:plb,unr:prd,hawk:npb, kle:npb} are explored with different fields after the discovery of wormhole \cite{Gs:1} as a solution of the Euclidean field equations invoking the axionic field in Einstein's gravity. In this context we specify the wormhole as a configuration governed by the Euclidean field equations, wherein two asymptotic regions of Robertson walker spacetimes are connected by a tube of minimum radius of scale factor $a(\tau_0)$ at some time $\tau_0$.  The axionic field in \cite{Gs:1} acts as an exotic field, which gives rise to violation of null energy condition. The wormhole solution demands violation of null energy condition \cite{visser:nec}. The violation can be minimised by introducing some geometric contribution instead of exotic fields in some theories, like the modified theory of gravity with higher order curvature terms\cite{capp:prd,harko:prd,lobo:fR}, the low energy heterotic string effective theory\cite{heter:string,met:string},
Kaluza Klein gravity\cite{sahdev:plb,wsson:AJ,Dzhu:prd}, etc.
\par
Physics onward GUT era is well understood and evolution of the universe from the inflationary era to the radiation dominated era, subsequent re-ionization, structure formation, etc are well known in the literature.
Physics of the pre-inflationary era, or the Planck era is not well understood in the literature and the chronological  scenario of subsequent physical configuration is not known; however, the wormhole, if it exits at the Planck era, subsequent cosmic evolution should reproduce a regime of inflation at later stage. Though the cosmological wormholes are considered \cite{Maeda:0901,cataldo:prd}, yet the viability of an inflationary era from an early wormhole configuration has not yet been clear in the literature. The evolution from an initial wormhole configuration at the Planck era to the subsequent inflationary scenario at later era may not be a smooth process, since the Planck era is determined by the quantum gravity, while the inflationary era can be described by the field theory and the Physics at these eras are completely different from the Planck era. Specifically, the term ``evolution" in cosmology is specified with respect to proper time $t$ and it is unique in the classical allowed domain, while the `` time" in the quantum field theory in the pre-inflationary era is specified by the Euclidean time $\tau$ and the corresponding background is the Euclidean space. The Lorentz time or the cosmic time ``$t$" in the classically allowed domain cannot probe the events or evolutionary scenario in the classical forbidden domain in general, as well as the time ``$\tau$" in the Euclidean domain cannot probe the Lorentz domain. So to probe the scenario outside its domain we need analytic continuation of the function from the Euclidean domain to the Lorentzian domain, vice verse (or the change of geometry) and the Wicks rotation $\tau=it$ serves its purpose. Thus to study evolution from an Euclidean wormhole to an inflationary scenario the transformation $\tau=it$ acts as an important tool.
\par
The work \cite{fuku:plb} is worthy to achieve an evolution of the Euclidean wormhole to a deSitter universe through a tube with oscillating radius invoking $R$, $R^2$ and $\Lambda$ terms in the action in Robertson Walker metric for closed three space section, however it depends on approximation. A reconstruction of the scalar tensor $f(R)$ function corresponding to evolving wormhole is considered in \cite{eur:sc}  an inhomogeneous spacetime generalizing Morris Thorn metric, while in \cite{TMP:darabi} Darabi presents the form of $f(\tilde{R})$ in Palatini $f(\tilde{R})$ cosmology as $f(\tilde{R})\sim \tilde{R}-c_1 \tilde{R}^2 +c$ for small $\tilde{R}$ from an Euclidean wormhole in $\mbox{O}^{\prime}$Hanlon theory. Though the works \cite{eur:sc,TMP:darabi} are reconstruction of $f(R)$ for evolving wormhole, however, they could not reproduce inflation from a wormhole configuration.
Interestingly, power law inflation \cite{ks:fR} is obtained asymptotically from an Euclidean wormhole in gravity with $R^{\frac{3}{2}}$ term using Noether symmetry. In a recent work, \cite{kk5:comm} it is shown that an Euclidean wormhole in the early universe evolved to an inflationary era away from the throat of the wormhole using $\tau=it$ in 5-dimensional Kaluza Klein spacetime minimally coupled with a scalar field.
\par
Our motivation is to study the viability of transition of an Euclidean wormhole configuration in the early era to the inflationary scenario invoking analytic continuation by $\tau=it$ in \textbf{$R^2$} gravity. It is important to note that all the gravitational theories contain at least linear $R$-term apart from higher order curvature contribution. However, pure $R^2$ gravity in the context of supergravity and superstring theory possess two important features \cite{koun:r2}, namely it is the only ghost free theory in quadratic curvature tensor and one can show it is a scale invariant gravitational theory. Under conformal transformation the  action of $R^2$ gravity is equivalent to a term linear in $R$ coupled minimally with a massless scalar field and a non-zero cosmological constant. Interestingly, traversable Lorentzian wormhole \cite{saza:ptp} solutions have been considered in the framework of $R^2$ gravity in time dependent inhomogeneous spacetime assuming constant Ricci scalar and trace-less energy momentum tensor. However, our work on wormhole is based on the Euclidean background of Robertson Walker metric.
 \par
We investigate the Euclidean field equations with a view to obtain possible wormhole solution in the very early era and consequence of them in the later epoch in Lorentz time $t$ using the modified theory of gravity in the spatially homogeneous and isotropic Euclidean background in 4-dimension. In a toy model, we consider $R^2$ term in the action since such term alone (without any matter field) admits inflationary solution \cite{a:plb} in cosmic time $t$. The field equations contains fourth derivative terms in the standard variational principle. Thus to get an alternative form of the field equations in second order we consider Boulware et al \cite{Bou:QT} proposal. In this proposal one can introduce a new variable, which is the first derivative of the action with respect to the highest derivative of the field variable that appears in the action after removing the total derivative terms in the action following the  works \cite{Bou:QT,hor:prd,aks:bm,bm:cqg,bm:05}. We consider the new variable with a view to obtain canonical variables and appropriate configuration space for canonical quantization to obtain the Wheeler deWitt equation to study the wave functional of the universe. The solution of the Wheeler deWitt equation yields the wave functional, which is oscillatory and finite in the Lorentzian background of Robertson Walker mini-superspace. However, the wave functional in the Euclidean background of Robertson Walker mini-superspace is finite at the classical singularity and dies away with expansion. Thus the quantum wormhole is possible following Hawking Page wormhole boundary condition in the Euclidean background of Robertson Walker mini-superspace. 
\par
We further present some solutions of the Euclidean field equations in $R^2$ gravity. We simplify the field equations introducing the idea of an extremum of the scale factor $a(\tau)$ at a given $\tau$ in the dynamics  through an integration constant in the combined field equation. Such an extremum of $a(\tau)$ is necessary to obtain possible wormhole solution, further the extremum of $a(\tau)$ at a given $\tau$ restricts the dynamics of spacetime geometry to a spacetime of constant curvature, which naturally leads to deSitter solution. Thereafter we consider piece-wise solutions depending on the contribution of different terms at different epoch for  $\kappa =0, \pm1$. Euclidean wormholes, as well as oscillatory solution of $a^2(\tau)$, are obtained as a function of Euclidean time $\tau$ for all 3-space curvature parameter $\kappa$. Oscillatory solution of $a^2(\tau)$ reduces to expanding solution with cosmic time $t$ under analytic continuation  $\tau=it$ and asymptotically it leads to an inflationary solution.
Important feature is to note from the piece-wise solutions that the universe beginning from an early Euclidean wormhole evolves to an oscillating universe at relatively greater $a(\tau)$ in $\tau$ and thereafter crossing the deSitter radius it transits to an inflationary scenario at later epoch using $\tau=it$ for $\kappa=1$.
Such type of transition from the Euclidean space to Lorentzian space are not allowed by the solutions for $\kappa=0,~-1$.
All the solutions satisfy the null energy condition.
\par
We present the field equations in the Lorentzian section in $R^2$ gravity in section 2. In section 3, we present the Wheeler deWitt equation and its solution, while in section 4 the solution for Euclidean wormholes are considered. We present a brief discussion on the null energy condition and appearance of spacetime of constant curvature in section 5 and finally we consider a brief conclusion in section 6.

\section{The Field equations in the Lorentzian theory in $R^2$ gravity}
To study wormhole configuration or quantum wormhole we first consider the action and the field equations. We consider canonical analysis in order to specify the appropriate configuration space and differential equation for the wave functional. Horowitz \cite{hor:prd} considered an action in the form
\begin{equation}\label{2.01}
S=-\frac{1}{4}\int d^4x \sqrt{-g
}[\alpha C_{\mu\nu\alpha\beta}C^{\mu\nu\alpha\beta} + \frac{2 \beta}{3} (R-4\Lambda)^2],
\end{equation}
where $C_{\mu\nu\alpha\beta}$ is the Weyl tensor, $R$ is the Ricci scalar, $\Lambda$ is the cosmological constant, $\alpha$ and $\beta$ are constants. In the weak energy limit the action \eqref{2.01}  reduces to the Einstein gravity. The field equations can be obtained in a simplified form introducing Robertson-Walker metric ansatz
\begin{equation}\label{2.1}
ds^2=-dt^2 + a^2(t)\Big[ \frac{dr^2}{1-\kappa r^2•} +r^2\big(d\theta^2 +\sin^2 \theta d\phi^2  \big)  \Big],
 \end{equation}
 where $a(t)$ is the scale factor and $\kappa$ is the three space curvature parameter and it can take $0,~\pm1$. The tensor $C_{\mu\nu\alpha\beta}$ vanishes trivially in the metric \eqref{2.1}, further assuming $\Lambda=0$ the action \eqref{2.01} retains only curvature squared term. The field equations gives rise to higher derivative terms, hence canonical formulation is not possible in the usual way. So we consider Boulware et al \cite{Bou:QT} proposal to reduce fourth order equations to an alternative second order formalism following \cite{Bou:QT,hor:prd,aks:bm,bm:cqg}.
The action \eqref{2.01} then leads to
\be\label{2.2}
S= -m \beta\int dt  \Big[ a \ddot{a}^2 + \frac{1}{a}(\dot{a}^2 +\kappa)^2 \Big]+S_{m_1},
\ee
where the Ricci scalar is $R= 6(\frac{\ddot{a}}{a}+\frac{\dot{a}^2}{a^2}+ \frac{\kappa}{a^2})$, an overhead dot stands derivative with respect to $t$ and $S_{m_{1}}=-m\beta \Big[ \int dt\frac{d}{dt}\Big\{2 \dot{a} (\kappa+\dot{a}^2)- \frac{4 }{3}\dot{a}^3\Big\}\Big]$ and  $m=12\pi^2$ is a constant. The integrand in \eqref{2.2} contains a term quadratic in $\ddot{a}$, so the Lagrangian cannot be obtained in terms of usual canonical variables. To overcome this problem and to reduce the fourth order gravity theory to a second order theory without changing degrees of freedom of the system we introduce new variable $Q$ following \cite{Bou:QT} as
\be\label{2.2a} m\beta~ Q= - \frac{\partial S}{\partial \ddot{a} },~ \mbox{which~ gives~~} Q=2 a\ddot{a}.
 \ee
Now replacing $ a\ddot{a}^2 $ by $Q$ using \eqref{2.2a} as
 $a \ddot{a}^2=\ddot{a}Q-\frac{Q^2}{4a} = \frac{d}{dt}(\dot{a}Q) - \dot{a}\dot{Q} -\frac{Q^2}{4a}$ in
equation \eqref{2.2} we can reduce the action following \cite{Bou:QT,aks:bm} as
\be\label{2.2a1}
S=m\beta\int d\tau \Big[  \dot{a}\dot{Q} - \frac{1}{a}(\dot{a}^2 +\kappa)^2 +{Q^2\over 4a}\Big]+ S_{m},
\ee
where $S_m= S_{m_{1}} - m\beta \int dt \frac{d}{dt}(\dot{a}Q)$ is the total surface term (or boundary term) and extraction of the surface term from the action is necessary to get well defined action principle. This term is analogous to the Gibbons-Hawking-York surface term of general relativity. Now with proper choice $S_m$ the Lagrangian in the Lorentzian space is
\begin{equation}
\label{2.2a2}
L_{_L}= M\Big[\dot{Q}\dot{a}  -\frac{(\dot{a}^2+ \kappa)^2}{a} +\frac{Q^2}{4a} \Big],
\end{equation}
where $M=m\beta$. The classical field equations can be obtained from the Lagrangian \eqref{2.2a2} in the usual way and they are second order equations in terms of the configuration space variable $\{a(t), Q(t)\}$. It is to be mentioned that the variation of $Q$ in \eqref{2.2a2} leads to definition of $Q$ given in \eqref{2.2a}.
The Hamilton constraint and a-variation equations are
\begin{equation}\label{2.2a21}
 \frac{\dot{a} \dot{Q} }{a^3}-\frac{Q^2}{4a^4}+ \frac{(\kappa+{\dot{a}}^2 )(\kappa-3{\dot{a}}^2)}{a^4} =0,
\end{equation}
\begin{equation}\label{2.2a22}
 {\ddot{Q} }-12 \frac{{\dot{a}}^2 {\ddot{a}}}{a•}- 4\kappa \frac{{\ddot{a}}}{a•}+ 4(\kappa+{\dot{a}}^2 )\frac{{\dot{a}}^2}{a^2•}-\frac{(\kappa+{\dot{a}}^2)^2•}{a^2•} +\frac{Q^2}{4a^2•}=0.
\end{equation}
Now using this $Q=2a\ddot{a}$ in \eqref{2.2a21} and \eqref{2.2a22} we have usual fourth order gravity theory in terms of $a(t)$, however the field equations in terms of $a(t)$ and $Q(t)$ represent the second order equations. Before considering Euclidean wormhole from the Euclidean field equations we consider possibility of quantum wormhole from the solution of the Wheeler deWitt equation. So, first we construct the Hamiltonian from the Lagrangian \eqref{2.2a2}.

\section{Quantum wormhole from the solution of Wheeler deWitt equation}
Quantum wormhole in the Hawking Page proposal is that the wave functional is finite at the vanishing scale factor and dies away at large three space geometry. So to study the wave functional of the universe we construct Wheeler deWitt equation in the background of Robertson Walker minisuperspace model for $\kappa=1$ in the $R^2$ gravity.
It is to mention that the wave functional $\Psi$ of the universe is a function of the configuration space. In order to specify appropriate configuration space and the Wheeler deWitt equation for $\Psi$, we consider canonical analysis. Now the canonical momenta conjugate to $a(t)$ and $Q(t)$ from the Lagrangian \eqref{2.2a2} are defined in the usual way
\begin{equation}
\label{2.2a3}
p_a= \frac{\partial L}{\partial\dot{a}}= M \Big[\dot{Q} - \frac{4\dot{a}(\dot{a}^2+1)}{a} \Big], ~~ p_Q=\frac{\partial L}{\partial\dot{Q}}=M \dot{a}
\end{equation}
and the field equations have already been given in \eqref{2.2a21} and \eqref{2.2a22}, however our main concern is to find the Hamiltonian, which turns to
\begin{equation}
\label{2.2a4}
{\cal{H}}=p_a \dot{a} + p_Q \dot{Q}-L_{_L}= \frac{p_a~p_Q}{M} +\frac{M}{a}(1 + \frac{{p_Q}^2}{M^2})- \frac{M Q^2}{4a}.
\end{equation}
Classical Hamiltonian is constrained to vanish, so quantum mechanically Hamiltonian operator annihilate the wave functional. Quantisation is performed \cite{hor:prd} in a representation of basic variables, where $a$ and $p_Q=M\dot{a}$ (where, $\dot{a}$ is related to the extrinsic curvature) are the configuration space variables.
Let us choose $\dot{a}=y$ and replace $p_Q$ by $M y$ using \eqref{2.2a3} and $Q$ by $\frac{p_y}{M}$ (as $p_y= \frac{\partial L_{_L}}{\partial\dot{y}}= 2Ma\dot{y}= M Q(t)$). So \eqref{2.2a4} reduces to
\begin{equation}
\label{2.2a5}
{\cal{H}}= y p_a - \frac{p_y^2}{4M a} +\frac{M}{a}(1 + y^2)^2.
\end{equation}
Now introducing $\hat{p_a}=-i\hbar \frac{\partial }{\partial a}$ and  $\hat{p_y}=-i\hbar \frac{\partial }{\partial y}$ in canonical quantisation, the Wheeler deWitt equation $\hat{{\cal{H}}}\Psi= 0  $ gives
\begin{equation}
\label{2.2a6}
 i\hbar a\frac{\partial \Psi}{\partial a} = \frac{\hbar^2}{4M y}\Big(\frac{\partial^2 \Psi}{\partial y^2} + \frac{s}{y} \frac{\partial \Psi}{\partial y}\Big)  +\frac{M}{y}(1 + y^2)^2\Psi,
\end{equation}
where $s$ is the operator ordering index. In \eqref{2.2a6} wave functional is $\Psi=\Psi[a(t), y(t)]$. The equation \eqref{2.2a6} has been obtained in an
 earlier work \cite{aks:bm} in quantisation of quadratic gravity. Though the calculations are identical, however our main concern is to study the possibility of quantum wormhole from the solution of Wheeler deWitt equation in the mini-superspace spanned by the variables $\{a, y\}$. The Wheeler deWitt equation in this form almost resembles with the Schrodinger equation, wherein the variable ``$\ln a(t)$" acts as the extrinsic time variable in quantum cosmology. Further, the continuity equation, hence the probability density can be obtained from \eqref{2.2a6} similar to the work \cite{aks:bm}. We consider solution of \eqref{2.2a6} instead of considering continuity equation, probability density, etc.  A trivial solution of \eqref{2.2a6} for $s=0$ is
\begin{equation}
\label{2.2a7}
\Psi[a, y]= \psi_0~ a~ \exp\Big[\frac{i2M}{\hbar}y(1-\frac{y^2}{3})  \Big],
\end{equation}
where the variables $a$ and $y$ in \eqref{2.2a7} are function of $t$ and we considered $y=\dot{a}$, so $y=a(t)H(t)$, where $H(t)= \frac{\dot{a}}{a}$ is the Hubble parameter. Thus the wave functional turns to
\begin{equation}
\label{2.2a8}
\Psi[a(t), y(t)]= \psi_0~ a(t)~ \exp\Big[\frac{i2M}{\hbar}a(t)H(t)\Big(1-\frac{a^2(t)H^2(t)}{3}\Big)  \Big].
\end{equation}
The wave functional $\Psi[a(t), y(t)]$ is an oscillatory function in general in the Lorentzian background of minisuperspace, though the wave functional is explicit independent of time $t$. It is important to note that the universe in the very early era at the Planck time is thought as the Euclidean space and the Euclidean time $\tau$ is appropriate to consider the configuration space variables, so we consider the wave functional in the Euclidean space background.
Thus under analytic continuation from the Lorentz space to the Euclidean space by $t=-i\tau$,  \eqref{2.2a8} leads to
\begin{equation}
\label{2.2a9}
\Psi[a(\tau), y(\tau)]= \psi_0~ a(\tau)~ \exp\Big[-\frac{2M}{\hbar}a(\tau)H(\tau)\Big(1+\frac{a^2(\tau)H^2(\tau)}{3}\Big)  \Big].
\end{equation}
The wave functional $ \Psi[a(\tau), y(\tau)] $ in the Euclidean background of minisuperspace is finite at the classical singularity and dies away with increasing $a(\tau)H(\tau)$, so the wave functional \eqref{2.2a9} supports the Hawking Page wormhole boundary condition in an expanding Euclidean universe. Hence quantum wormhole exists in $R^2$ gravity in the Euclidean background of minisuperspace. Now we consider the solution of the Euclidean field equations to study possible of Euclidean wormhole.

\section{Solution of the Euclidean field equations to study possible wormhole configuration}
In study of Euclidean wormhole we first consider the Euclidean field equations. The field equations can be obtained from the Lorentzian field equations \eqref{2.2a21} and \eqref{2.2a22}  using analytic continuation $t \longrightarrow \tau=it$, which are given as
\be\label{2.3}
 {a'} {Q'}+\frac{Q^2}{4a}- \frac{(\kappa-{a'}^2 )(\kappa+3{a'}^2)}{a} =0,
\ee

\be\label{2.4}
 {Q''}+12 \frac{{a'}^2 {a''}}{a•}- 4\kappa \frac{{a''}}{a•}+ 4(\kappa-{a'}^2 )\frac{{a'}^2}{a^2•}+\frac{(\kappa-{a'}^2)^2•}{a^2•} -\frac{Q^2}{4a^2•}=0,
 \end{equation}
where $ Q(\tau)=-2a(\tau) a''(\tau)$ from the definition of $Q$ in \eqref{2.2a} in the Euclidean space. Solution of the field equations needs special care as they are coupled non linear equations. Evolution of wormhole is considered in \cite{fuku:plb} assuming
$a'^2= 1+f(a)$  in \eqref{2.3} for $\kappa=1$, where $f(a)$ is arbitrary function of $a$ and eventually they \cite{fuku:plb} extract solution when $(f(a)-f_0)$ is very small, where $f(a)=f_0$ at the extreme regime of $a$. In our presentation, we consider exact solution of evolving wormhole using all the field equations in $R^2$ gravity. It is important to mention that a wormhole has two asymptotic regions connected by a tube with a minimum  radius, so $a(\tau)$ should have $a^{\prime}=0$ and $ a^{\prime\prime}>0$ at some $\tau=\tau_0$ at the minimum, otherwise $a(\tau)$ is finite.
 Now multiplying \eqref{2.4} by $a$ and taking sum of it with \eqref{2.3} gives

\be\label{2.5}
\Big( aQ'+4a'^3 -4\kappa a'\Big)'=0.
\ee
The solution of $a(\tau)$ from  \eqref{2.5} is not possible in closed form. Thus considering $Q=-2aa''$ in \eqref{2.5} we have

\be\label{2.6}
aQ'+4a'^3 -4\kappa a'=C=  -2a'\Big[a^2 \dfrac{d}{da•}(a'')+ aa''-2a'^2+2\kappa \Big],
\ee
where $C$ is an integration constant, further \eqref{2.6} can be  simplified as

\be\label{2.7}
(a'^2)_{xx}-4 a'^2 +4\kappa= -{C\over a'},
\ee
where $x= \ln a$ and subscript $x$ denotes derivative with respect to $x$. The constants are usually determined by the initial condition on the variable and its derivative, however, one may have a prior information depending on the configuration of the physical system. Since we are seeking wormhole configuration, one should have $a'=0$ and $a'' >0$ at some $\tau$ for existence of non-vanishing minimum of  $a(\tau)$ at the throat of the wormhole, so equation \eqref{2.6} yields $C=0$ without loss of generality. So integration of \eqref{2.7} for $C=0$ yields

\be\label{2.8}
 a'^2=\kappa + c~ a^2 + \frac{c_1}{a^2},
 \ee
where $c$ and $c_1$ are constants. It is important to note that $\frac{c_1}{a^2}$ term in \eqref{2.8} is dominant in the very early universe, while the term $c a^2$ has a large contribution in the late epoch when $a(\tau)$ is sufficiently large. Further the sign of $c$ and $c_1$ are also important in determining the solution. So the different terms lead to distinct scenario at different epoch in cosmic evolution. In wormhole the radius $a_0$ at the throat at time
$\tau=\tau_0$ from \eqref{2.8} satisfies
\be\label{2.9}
a'(\tau_0)=0 \rightarrow c~ a_0^4 + \kappa a_0^2 +  c_1=0, ~\hbox{wherein}~~a_0''=\frac{\kappa}{a_0}+2ca_0=-\frac{\kappa}{a_0}-\frac{2c_1}{a_0^3} =\frac{\sqrt{\kappa^2-4cc_1}}{a_0}>0~~
 \ee
for minimum of $a(\tau)$.
\par
The equation \eqref{2.8} allows class of solutions depending on the sign of $\kappa$, $c$ and $c_1$. In the forthcoming section, we classify them according to the values of $\kappa$; further, we specify the details of the solutions depending on $c$ and $c_1$. To get a clear view and difference among the solutions depending on different signs of above constants we invoke separate symbols for $c$ and $c_1$ in each case. So we consider the piece-wise solutions of \eqref{2.8} according to the weightage of individual terms in the right hand side of \eqref{2.8}. Finally, we consider interpretation of the solutions relevant to cosmic evolution.

\subsection{Possible solutions and cosmic scenario for $\kappa=0$:}
Now we consider solution of \eqref{2.8} with restriction on $c$ and $c_1$ for $\kappa=0$.

\subsubsection{Solution with $c_1 <0$ and $c>0$ for $\kappa=0$:}
The term $\frac{c_1}{a^2}$ in \eqref{2.8} dominates in the very early universe, however the term $\frac{c_1}{a^2}$ alone does not yield wormhole solution for $\kappa=0$. A wormhole configuration from \eqref{2.8} is allowed for $\kappa=0$ when the constants are restricted to satisfy $c_1=-H_0^2 <0$ and $c=h_0^2>0$. Assuming the extremum of $a(\tau)$ (i.e. $a'=0$) at $\tau=0$ the scale factor turns to
\begin{equation}\label{2.10}
a^2(\tau) = a^2_0 \cosh(2h_0\tau),
\end{equation}
which is a wormhole solution, where radius of the wormhole at the throat is $a_0= \sqrt{\frac{H_0}{h_0}}$. Now analytic continuation with $\tau=it$ yields an oscillating $a(t)$ for all $t$, which is not acceptable in Lorenzian spacetime.
\subsubsection{Solution with $c_1 >0$ and $c<0$ for $\kappa=0$:}
The scale factor $ a(\tau) $ satisfying restriction
 $c_1 =H_1^2>0$ and $c=-h_1^2 <0$ leads to
 \begin{equation}\label{2.10a} a_0=  \sqrt{\frac{H_1}{h_1}} ~~\hbox{and} ~~a_0''= -2h_1^2a_0
\end{equation}
from \eqref{2.9}, so the extremum of $a(\tau)$ at $\tau_0$ does not yield a lower bound, rather it gives a upper bound. Thus wormhole configuration is not possible with $c_1 >0$ and $c<0$.
However, an inflationary solution may be obtained in the Lorentzian segment using $\tau=it$ in \eqref{2.8} with $c_1 >0$ and $c<0$ as

\be\label{2.11}
a^2(t)= \frac{r_1}{2}~ e^{2h_1t} + \frac{H_1^2}{2r_1h_1^2}e^{-2 h_1t},
\ee
where $r_1$ is a constant. So, the solution shows an exponential expanding era asymptotically at $2h_1 t>>1$ and the inflationary solution would not emanate from an early Euclidean wormhole for $\kappa=0$.\\
Thus from above solutions for $\kappa=0$  inflationary era cannot be obtained from an early Euclidean wormhole in considering transition from an Euclidean space to the Lorentzian space with a specific choice (or unique choice) of $c$ and $c_1$.
\subsection{ Possible solutions and cosmic scenario for $\kappa=-1$: }
Now we consider possible solutions of \eqref{2.8} with a few restriction on $c$ and $c_1$ for $\kappa=-1$.

\subsubsection{Solution when $\frac{c_1}{a^2} >> ca^2$ for $c_1>0$ and $\kappa=-1$:}
The term $\frac{c_1}{a^2}$ in \eqref{2.8} is dominating compared to the term $c a^2$ in the very early universe, thus neglecting $c a^2$ in \eqref{2.8} the extrema of $a(\tau)$ at some $\tau_0$ leads $a_0=H_1$, wherein $a_0''=-\frac{1}{H_1}$ for $c_1=H_1^2$. So wormhole is not possible in this case, rather this extremum represents an upper bound with respect to $\tau$. However in the Lorentzian section (using $\tau=it$) $H_1$ represents a lower bound, and the scale factor $a(t)$ with $t$ from \eqref{2.8} is
\be\label{2.12}
a^2(t)= H_1^2+t^2.
\ee
 The solution \eqref{2.12} is symmetric with $t$ and the radius of the universe at the minimum is $ H_1$ at $t=0$. Further $a(t) \approx t$ at $t>> H_1$, so we have a power law expansion.
Further the effect of $c a^2$ term in \eqref{2.8} will be more pronounced with expansion and one should consider the contribution of other terms in \eqref{2.8} at later epoch.

\subsubsection{Solution with $c< 0$ and $ c_1 >0$ for $\kappa=-1$:}
The extremum of $a(\tau)$ from \eqref{2.8} with restriction $c=-h_1^2 <0$ and $c_1=H_1^2> 0$ gives $a_0''<0$ and $a_0^2=[-1+\sqrt{1+4h_1^2H_1^2}~]/2h_1^2$ at the extremum, so $a_0$ leads to a maxima with $\tau$, however it reduces to a non-vanishing minimum with $t$ using $\tau=it$. The scale factor from \eqref{2.8} with these $c$ and $c_1$ leads to
\be\label{2.12a}
a^2(t)= \frac{1}{2h_1^2}\Big[-1+ \frac{r_2}{2}e^{2h_1t}+ \frac{2h_1^2H_1^2}{r_2}e^{-2h_1t}  \Big],
\ee
where $r_2$ is constant and a comparison of above $a_0^2$ with the
expression from \eqref{2.12a} (assuming $t=0$) gives \\$r_2=\pm1 + \sqrt{1+4h_1^2 H_1^2}$. The scale factor \eqref{2.12a}  gives an exponential expansion in the asymptotic regime $ 2h_1t>> 1 $.
\subsubsection{Solution with $c>0$ and $4c c_1 < 1$ for $\kappa=-1$:}
The equation \eqref{2.8} allows a minimum of $a(\tau)$ with restriction $c=h_0^2>0$, which  gives $ a_0^2=(1+\sqrt{1-4h_0^2 c_1}~)/2h_0^2$ with $a_0''>0$ for $4h_0^2 c_{1}<1$. The scale factor from \eqref{2.8} then reduces to
\be\label{2.12a1}
a^2(\tau)= \frac{1}{2h_0^2}\Big[1+r_3 \cosh(2h_0\tau)  \Big],
\ee
where $r_3= \sqrt{1-4h_0^2 c_1}<1$ assuming minimum of $a(\tau)$ is at $\tau=0$. Thus \eqref{2.12a1} gives a wormhole with a radius $a_0= \sqrt{\frac{1+r_3}{2h_0^2}}$ at the throat. The scale factor is quite large at greater $\tau$ and $a(\tau)$ increases exponentially with $\tau$ at $2h_0 \tau >>1$, eventually the term $ca^2$ in \eqref{2.8} dominates over $\frac{c_1}{a^2}$ at later era. Further
analytic continuation of the solution with $\tau=it$ yields an oscillating universe in time $t$ with the period $\frac{1}{2 h_0}$, where non-vanishing lower bound and upper maximum of $a(t)$ are respectively $ \sqrt{\frac{1-r_3}{2h_0^2}}$  and $ \sqrt{\frac{1+r_3}{2h_0^2}}$, provided $r_3<1$. Thus an oscillatory universe in time $``t"$ in the Lorentzian spacetme is equivalent with an Euclidean wormhole with choice $ 4h_0^2c_1<1 $ for $\kappa=-1$.

\subsubsection{Solution when $ca^2>>\frac{c_1}{a^2}$ for $c>0$ and $\kappa=-1$:}
The solution of \eqref{2.8} when the scale factor is sufficiently large, or with the condition $ca^2>>\frac{c_1}{a^2} $ in \eqref{2.8} for $\kappa=-1$ gives

\be\label{2.12b}
 a(\tau)= \frac{r_4}{2h_0^2}e^{h_0\tau} + \frac{1}{2r_4•}e^{-h_0\tau},
\ee
where $r_4$ is constant and $c=h_0^2$. Now the extremum of $a(\tau)$ in \eqref{2.12b} at $\tau=0$ gives $r_4= h_0$. So \eqref{2.12b} leads to
\be\label{2.12c}
a(\tau) = \frac{1}{h_0}\cosh(h_0\tau),
\ee
which is also a wormhole solution. So all the solutions \eqref{2.12a1}-\eqref{2.12c} for $\kappa=-1$ yield Euclidean wormhole configuration. Above piecewise solutions for $\kappa=-1$ show that none of the Euclidean wormhole evolves to an exponential expanding era with $t$ under analytic continuation $\tau=it$ for a specific choice of $c$ and $c_1$, rather with $c_1>0$ and $c>0$ the early universe shows a wormhole configuration \eqref{2.12} in $t$ and in the later epoch with greater scale factor the universe evolves to the  Euclidean wormholes \eqref{2.12a1}-\eqref{2.12c}.

\subsection{ Possible solutions and cosmic scenario for $\kappa=1:$}
In this section we consider all possible solutions of \eqref{2.8}  with a few restriction on $c$ and $c_1$ for $\kappa=1$.

\subsubsection{ Solution under $\frac{c_1}{a^2} >> ca^2$ with $c_1 <0$ for $\kappa=1$:}
In the early universe when $a(\tau)$ is very small the contribution of $c a^2$ term is not significant compared to $\frac{c_1}{a^2}$ in \eqref{2.8}, so we can neglect the contribution of $c a^2$ in the early epoch of evolution. Then \eqref{2.8} with $c_1=-H_0^2<0$ leads to
\be\label{2.14}
a^2(\tau)= \tau^2+ H_0^2,
\ee
where $a_0''=\frac{1}{H_0}$ and $a_0=H_0$ at $\tau=0$, so it represents a wormhole solution and the radius at the throat of the wormhole is $H_0$. The scale factor is symmetric with $\tau$ and it is increasing with $\tau$. Now at sufficiently large scale factor the contribution of $c a^2$ also dominates in \eqref{2.8}. However, at some intermediate stage the contribution of all the terms in \eqref{2.8} play an important role in evolution.
\subsubsection{Solution with $c>0$ and $c_1 <0$ for $\kappa=1$:}
The equation \eqref{2.8} with restrictions  $c=h_0^2>0$ and $c_1=-H_0^2 <0$ yields a minimum of $a(\tau)$ for $ \sqrt{1+4h_0^2H_0^2}> 1$ and the value of $a(\tau)$ at the minimum  is $a_0= \sqrt{\frac{r_5-1}{2h_0^2}}$, where $r_5= \sqrt{1+4h_0^2H_0^2}$. The scale factor from \eqref{2.8} gives
\be\label{2.14a}
a^2(\tau)= \frac{1}{2h_0^2} \big[-1 +r_5 \cosh(2 h_0 \tau) \big],
\ee
so we have a wormhole solution for $c>0$ and $c_1 <0$. Again the scale factor is large enough with $\tau$ and $a(\tau)$ increases exponentially with $\tau$ at $2h_0 \tau >>1$. However, at small $\tau$, \eqref{2.14a} leads $a^2(\tau)\approx\frac{r_5-1}{2h_0^2}+ r_5\tau^2$ for $2h_0\tau<1$ with lowest order contribution, which agrees with the scenario of the early universe given in \eqref{2.14} with re-definition of $\tau$. At large scale factor the contribution $\frac{c_1}{a^2}$ is very small and $ca^2$ plays a crucial role in evolution and dynamics will be different with reverse sign of $c$. So we present another solution with different sign of $c$, but with $c_1<0$.

\subsubsection{Solution with $c<0$ and  $c_1 <0$ for $\kappa=1$:}
The extremum of $a(\tau)$ from \eqref{2.8} with $c<0$ and $c_1 <0$ leads a maximum of $a(\tau)$ with the value $a_0^2=\frac{1}{2h_1^2} (1 + r_6~)$ and $a_0''<0$ in the Euclidean space, where $r_6 = \sqrt{1-4h_1^2 H_0^2} < 1$ ( or $4h_1^2H_0^2<1$) with $c=-h_1^2<0$ and $c_1=-H_0^2<0$. So wormhole solution is not possible with $c<0$ and $c_1<0$, rather \eqref{2.8} leads to an oscillatory solution with $\tau$, which reads as
\be\label{2.14c}
a^2(\tau)= \frac{1}{2h_1^2} \big[1 + r_6 \cos(2 h_1 \tau) \big].
\ee
The solution \eqref{2.14c} leads to an oscillating universe in $\tau$ with a period $\frac{1}{2 h_1}$. Now introducing analytic continuation $\tau=it$ in \eqref{2.14c} we get
\be\label{2.14c1}
a^2(t)= \frac{1}{2h_1^2} \big[1 +r_6 \cosh(2 h_1 t) \big].
\ee
The expression of $a(t)$ by \eqref{2.14c1} represents a non-vanishing minimum of $\sqrt{\frac{r_6 +1}{2h_1^2}}$ at $t=0$ and $a(t)$ is symmetric with $t$ in the Lorentzian space. The solution also leads to an exponential expanding universe with $t$ at an era $2 h_1 t>>1$. So an oscillating solution \eqref{2.14c} in Euclidean time $\tau$ is equivalent with an expanding solution \eqref{2.14c1} in $t$ the Lorentzian space.
\subsubsection{ Solution with $ca^2 >>\frac{c_1}{a^2} $ and $c<0$  for $\kappa=1$:}
In course of evolution when the scale factor is sufficiently large at later epoch of $\tau$, we can assume $ \frac{c_1}{a^2} \simeq 0$.
The scale factor from \eqref{2.8} then reduces to
\be\label{2.15}
a^2(\tau)= \frac{1}{h_1^2} \sin^2\big(h_1\tau\big),
\ee
apart from integration constant, where $c=-h_1^2<0$. The solution of the scale factor is valid as long as $a(\tau)h_1 < 1$, which represents an oscillating universe with $\tau$ within a maxima $1/h_1 $ (deSitter radius) and vanishing minima radii. Now in the era when $a(\tau) h_1> 1$ the solution from \eqref{2.8} reduces to
\be\label{2.16}
a(\tau)= \frac{1}{h_1} \cosh \big(ih_1\tau\big),~~ \hbox{and it reduces to}~~a(t) = \frac{1}{h_1} cosh\big(h_1t\big),
\ee
under analytic continuation with $\tau=it$, which yields an expanding universe with the Lorentz time. Further in the asymptotic domain $t>> \frac{1}{h_1}$, the universe expands exponentially with $t$. In this case at large $t$ we have an exponential expansion. The universe before crossing deSitter radius is oscillating with Euclidean time $\tau$ and after crossing the deSitter radius it enters to an era of exponential expansion. This agrees with the idea of \cite{fuku:plb}, however, our results are independent of approximation.
Thus to sum up the results for $\kappa=1$ we have a configuration of Euclidean wormhole \eqref{2.14} in the very early universe, eventually the cosmic scenario evolves through \eqref{2.14c}-\eqref{2.15} to an inflationary era \eqref{2.16} in the asymptotic regime of time $t$ using $\tau=it$ with choice $c_1<0$ and $c<0$.
\par
In a nutshell, we can sum up the results of the piece-wise solutions saying that the transition from an early Euclidean wormhole to the inflationary era in the Lorentzian section is not possible for $\kappa=0$ and $\kappa=-1$ using $\tau=it$.  However, we can interpret and bring together the piece-wise solutions of \eqref{2.8} for $\kappa=1$  representing the cosmic scenario of the universe. Beginning with the Euclidean wormhole solution \eqref{2.14} in the very early era,  the subsequent cosmic evolution passes through an era and approaches to an oscillating universe with $\tau$ in solution \eqref{2.14c} and then \eqref{2.15}. Finally, the universe after crossing the deSitter radius enters to an era of faster expansion, given by \eqref{2.16} and asymptotically one can realize an exponential expansion with Lorentz time. So the evolution from an Euclidean wormhole to an exponential expanding era can be achieved only for $\kappa=1$ by analytic continuation $\tau=it$, which is in sharp contrast with the scenario for $\kappa=-1$ and $\kappa=0$.
In this context an interesting case appears, wherein an Euclidean wormhole \eqref{2.14} in the very early universe could evolve to an inflationary solution \eqref{2.16} with $t$ at large scale factor only for $\kappa=1$.
\section{Background geometry as a consequence of wormhole in $R^2$ gravity and Null energy condition:}
In  general relativity, wormholes are supported by the exotic matter, which leads to a stress energy tensor $T_{\mu\nu}$  that violates null energy condition (NEC). The NEC is given by $T_{\mu\nu}k^\mu k^\nu  \geq 0$, where $k^\mu$ is any null vector.  The violation of NEC can be minimised by introducing some geometric contribution \cite{capp:prd,harko:prd,lobo:fR,heter:string,met:string}, etc.
instead of exotic field.  In our study of $R^2$ gravity, no such energy momentum tensor exists,  however the condition $R_{\mu\nu}k^\mu k^\nu  \geq 0 $  acts as an alternative to the NEC from the Raychaudhuri equation. To evaluate it we first present an important result relevant to wormhole solution and spacetime geometry. The field equation in $R^2$ gravity is
\be\label{2.17}
R R_{\mu\nu}-\frac{R^2}{4}g_{\mu\nu}-R_{;\nu;\mu} + R^{; \alpha}_{~;\alpha}~ g_{\mu\nu}=0,
\ee
whose trace is $ R^{; \alpha}_{~;\alpha}=0 $ and it leads to $ \dot{R}a^3=R_0=$constant. Now using \eqref{2.8} with $a'=-i~\dot{a}$ and the Ricci scalar $R= 6(\frac{\ddot{a}}{a}+\frac{\dot{a}^2}{a^2}+ \frac{\kappa}{a^2})$  in $\dot{R}a^3=R_0$, we get $R_0=0$, which shows that the Ricci scalar $R$ is a constant with $t$. So the spacetime satisfies
\be\label{2.18}
 R_{\mu\nu}=\frac{R}{4}g_{\mu\nu},
\ee
which is a condition for spacetime of constant curvature and in general it allows deSitter spacetime. Kehagias et al \cite{keha:r2} also obtained this category of spacetime in $R^2$ gravity in different context. It is to be mentioned that $R^2$ gravity does not allow a spacetime of constant curvature in general, however the use of \eqref{2.8}, which governs the dynamics pertaining to extremum of $a(\tau)$ leads to the spacetime of constant curvature. So inclusion of an extremum in the scale factor at a given era, which is necessary in the Euclidean wormhole leads to a spacetime of constant curvature in $R^2$ gravity.  Now for any null vector $k^\mu$ the equation \eqref{2.18} gives $R_{\mu\nu}k^\mu k^\nu =0$. So above wormhole solutions satisfy NEC, hence the spacetime is not null geodesically complete, i.e.,  we cannot extend null lines arbitrarily and focusing of null geodesics are possible. From this discussion we may state that the existence of minimum in wormhole geometry (which is obvious) leads to deSitter solution in $R^2$ gravity.

\section{Discussion}
Physics of the pre-inflationary universe is not well understood, as the regime is beyond the realm of Einstein's gravity, rather it is determined by the quantum gravity. The initial cosmic singularity, or other relevant configuration of the universe or the wormhole in the Planck era is thus determined by the quantum theory, wherein the spacetime is assumed to be Euclidean. However, the field theory in the Hyperbolic spacetime with time $t$ can explain the inflationary scenario at the GUT era. Thus to explore Physics one may seek viability of evolution from a wormhole configuration to the inflationary scenario. So we consider a toy model of $R^2$ gravity to study such evolution or possible transition. We consider $R^2$ term since it has a huge contribution in the very early universe as well as it reveals inflation.

\par
We simplify the fourth order Euclidean field equations to a second order field equations introducing a new variable with a view to get canonical variables following \cite{Bou:QT,aks:bm} proposal. We consider canonical analysis to specify appropriate configuration space required for the Wheeler deWitt
equation to study the wave functional of the universe. The solution of the Wheeler deWitt equation reveals that the wave functional satisfies Hawking Page wormhole boundary condition ( necessary for quantum wormhole ) in the Euclidean background of Robertson Walker space with $\tau=it$, while in the Lorentzian space the wave functional is oscillatory.
\par
Further, in consideration of wormhole from the Euclidean field equations in the background of Robertson Walker space in $R^2$ gravity, we simplify the field equations introducing the idea of extremum (which is minimum for wormhole) of $a(\tau)$ through a constant $C$ in \eqref{2.7} (the constant $C=0$ for existence of extremum of $a(\tau)$ ), thereafter we consider the solutions of the simplified equation \eqref{2.8} for possible wormhole configuration or other cosmic scenario. The idea of extremum of $a(\tau)$ at some $\tau$ reduces the dynamics of spacetime to a spacetime of constant curvature, which naturally yields deSitter solution. The solution of above mentioned equation allows Euclidean wormhole as well as oscillatory solution of $a^2(\tau)$ depending on the constants and the curvature parameter $\kappa$. However, the oscillatory solution of $a^2(\tau)$ reduces to a expanding solution with $\tau=it$, which asymptotically yields exponential expansion with time $t$.
We can summarise the piece-wise solutions for $\kappa=0$ and $\kappa=-1$ saying that the transition from an early Euclidean Wormhole to an inflationary era asymptotically is not possible under analytic continuation $\tau=it$. However, the piece-wise solutions can be brought together to trace the evolution in chronological sequence depending on the relative values of $a(\tau)$ for $\kappa=1$. The cosmic evolution beginning with an Euclidean wormhole \eqref{2.14} in very early era approaches to an oscillating universe with $\tau$ in solution \eqref{2.14c} and then \eqref{2.15}. Continuing evolution the universe after crossing the deSitter radius enters to an era of faster expansion \eqref{2.16} and asymptotically one can obtain exponential expansion with $t$ using $\tau=it$. The new feature of our solution is that we can achieve
exponential expansion naturally in the later epoch beginning from a Euclidean wormhole in the very early universe for $\kappa=1$.
\par
Finally, we conclude that the quantum wormhole for $\kappa=1$ exists in the  Euclidean background of Robertson Walker mini-superspace from the solution of the Wheeler deWitt equation, while the wave functional is finite and oscillatory in the Lorentzian background.
In addition, exponential expansion of the universe can be obtained naturally in the later epoch beginning from an early Euclidean wormhole for $\kappa=1$. Thus introduction of wormhole configuration in the very early universe is well justified with onset of inflation at later epoch.

\end{document}